\documentclass[11pt,showpacs,preprintnumbers,amsmath,
amssymb,
aps,prd,nofootinbib, 
superscriptaddress
]{revtex4}

\usepackage{epsfig}
\usepackage{hyperref}
\usepackage{graphicx,epsf}
\usepackage{color}
\usepackage{bm}
\usepackage{psfrag}

\def\be{\begin{equation}}
\def\ee{\end{equation}}
\def\beb{\begin{equation*}}
\def\eeb{\end{equation*}}
\def\bea{\begin{eqnarray}}
\def\eea{\end{eqnarray}}
\def\beab{\begin{eqnarray*}}
\def\eeab{\end{eqnarray*}}

\def\bi{\begin{itemize}}
\def\ei{\end{itemize}}

\def\cs2{c_{\rm{s}}^2}

\def \beg {\begin{enumerate}}
\def \en {\end{enumerate}}

\def\cs{c_{\rm{s}}^2}

\begin{document}
\title{New astrophysical bounds on ultralight axionlike particles}

\author{Nilanjan Banik}
\affiliation{Department of Physics, University of Florida, 
Gainesville, Florida 32611, USA}
\affiliation{Fermi National Accelerator Laboratory, Batavia, Illinois 60510, USA}
\author{Adam J.~Christopherson}
\affiliation{Department of Physics, University of Florida, Gainesville, 
Florida 32611, USA}
\author{Pierre Sikivie}
\affiliation{Department of Physics, University of Florida, Gainesville, 
Florida 32611, USA}
\author{Elisa Maria Todarello}
\affiliation{Department of Physics, University of Florida, Gainesville, 
Florida 32611, USA}

\date{\today}

\begin{abstract} 

Motivated by tension between the predictions of ordinary cold dark matter (CDM) 
and observations at galactic scales, ultralight axionlike particles (ULALPs) 
with mass of the order $10^{-22}~{\rm eV}$ have been proposed as an alternative 
CDM candidate.  We consider cold and collisionless ULALPs produced in the early 
Universe by the vacuum realignment mechanism and constituting most of CDM. The 
ULALP fluid is commonly described by classical field equations. However, we show 
that, like QCD axions, the ULALPs thermalize by gravitational self-interactions and 
form a Bose-Einstein condensate, a quantum phenomenon. ULALPs, like QCD axions, 
explain the observational evidence for caustic rings of dark matter because they 
thermalize and go to the lowest energy state available to them. This is one of 
rigid rotation on the turnaround sphere. By studying the heating effect of 
infalling ULALPs on galactic disk stars and the thickness of the nearby 
caustic ring as observed from a triangular feature in the IRAS map of our 
galactic disk, we obtain lower-mass bounds on the ULALP mass of order 
$10^{-23}$ and $10^{-20}~{\rm eV}$, respectively.   

\end{abstract}


\maketitle

\section{Introduction}
\label{sec:intro}

Although the existence of dark matter is very well supported by data collected 
from various sources, there is no general agreement at present on what constitutes 
the dark matter. Two major candidates are the class of weakly interacting massive 
particles, and the QCD axion. The latter is a favored extension of the 
standard model of particle physics that solves the strong \textit{CP} problem \cite{Peccei1977, 
Weinberg1978, Wilczek1978} in addition to being a viable dark matter candidate 
\cite{Preskill1983, Abbott1983, Dine1983}.

Ultralight axionlike particles (ULALPs) with a wide range of masses between 
$10^{-33} \, {\rm eV} \leq m \leq 10^{-18}\, {\rm eV}$, the so-called ``Axiverse", 
\cite{Svrcek:2006yi, Arvanitaki:2009fg, Ringwald:2012hr}, are predicted in string 
theory-based extensions of the standard model. They are dark matter candidates 
as well, with properties similar to the QCD axion but much lighter.  If 
sufficiently light, they suppress structure formation on small scales 
\cite{Press,Sin,Goodman,Hu2000,Amendola:2005ad,BoylanKolchin:2011de, 
Shapiro,Park:2012ru} because they have a Jeans length \cite{Khlopov,Bianchi},
\be
\ell_J = (16 \pi G \rho m^2)^{-{1 \over 4}} =
1.01 \times 10^{14}~{\rm cm}
\left({10^{-5}~{\rm eV} \over m}\right)^{1 \over 2}                  
\left({10^{-29}~{\rm g/cm}^3 \over \rho}\right)^{1 \over 4}
\label{Jeans}
\ee
where $\rho$ is the ULALP mass density.  

Numerous bounds have been placed on the ULALP mass using observational 
data.  V. Lora \textit{et al}. \cite{Lora} found that the mass range 
$0.3 \times 10^{-22} < m < 10^{-22}$ eV provides a best fit 
to the properties of dwarf spheroidal galaxies. The cosmic 
microwave background anisotropy observations require 
$m > 10^{-24}~{\rm eV}$ \cite{Hlozek:2014lca}. Numerical 
simulations of structure formation with ULALP dark matter 
have been carried out and found to give a good description 
of the core properties of dwarf galaxies when 
$m \sim 10^{-22}$ eV \cite{Nature,Schive:2015kza}.   
ULALP dark matter with mass $m \sim 10^{-21}$ eV was
found to alleviate the problems of excess small-scale 
structure that plague ordinary cold dark matter \cite{MS}. 
Recent work finds that data on the Draco II and Triangulum II 
dwarf galaxies are best fit by a ULALP with mass 
$m \sim 3.7 - 5.6 \times 10^{-22}~{\rm eV}$ \cite{Calabrese:2016aa},
while other authors obtain $m < 0.4 \times 10^{-22}$ eV from Fornax 
and Sculptor data \cite{Gonzales}.  Because ULALPs cause reionization 
to occur at a lower redshift, it has been argued that reconstruction 
of the UV-luminosity function restricts 
$m \gtrsim 10^{-22} \,{\rm eV}$ \cite{Bozek2015}.

L. Hui \textit{et al}. \cite{Hui} have recently written a comprehensive overview of 
the ULALP literature, including a discussion of the relaxation of ULALP dark 
matter in gravitationally bound objects. Relaxation (also known as thermalization)
is the key ingredient for Bose-Einstein condensation to occur. Ref.\cite{bhupal} have studied the effects of Bose-Einstein condensate of 
ultralight dark matter on the propagation of gravitational waves.  We show in 
Sec. III how the analysis of Ref. \cite{Hui} relates to the earlier 
results in Refs. \cite{CABEC,Erken,Saikawa,Berges}.

A relic ULALP population can be produced thermally (generated from the 
radiation bath) or nonthermally by the vacuum realignment mechanism. The 
former class of ULALPs behaves like dark radiation. In this paper, we are 
solely interested in ULALPs generated by the vacuum realignment mechanism, 
which behave like cold dark matter.

In Sec. II, we show that ULALPs thermalize via gravitational self-interactions 
and hence form a Bose-Einstein condensate, in a manner analogous to QCD axions. This is something that 
the previous works have not taken into account. In Sec.III, we estimate a lower 
mass bound on ULALPs by requiring that the infalling ULALPs do not excessively heat 
the galactic disk stars. In Sec. IV, we obtain a stronger bound from the sharpness 
of the fifth caustic ring of the Milky Way galaxy which appears as a triangular 
feature in infrared astronomical satellite (IRAS) and Planck maps.  Sec. V provides a summary.

\section{Bose-Einstein condensation of ULALPs}

Bose-Einstein condensation occurs if the following conditions are 
satisfied:  i) the system is composed of a huge number of identical 
bosons, ii) these particles are highly degenerate, iii) their 
number is conserved, and iv) they thermalize.  Bose-Einstein
condensation means that most of the particles go to the lowest 
energy state available through the thermalizing interactions.  
We show below that the cold ULALP fluid produced by the vacuum 
realignment mechanism satisfies all four conditions.

The ULALP is described by two parameters: its mass $m$, which 
sets the time when the ULALP field begins to oscillate in the 
early Universe and its decay constant $f$, with dimensions of 
energy, which sets the magnitude of its initial misalignment and 
the strength of its interactions.  Unlike the case of QCD axions, 
$m$ and $f$ are independent parameters.  Also, unlike the case of 
QCD axions, the ULALP mass is taken to be temperature independent. 
The time when the ULALP field starts to oscillate is of order 
$t_1 \equiv 1/m$.  The ULALPs produced by the vacuum realignment 
mechanism \cite{Preskill1983, Abbott1983, Dine1983} have at that 
time number density $n(t_1)\sim m \phi_1^2$, where $\phi_1$ is the 
value of the field then.  $\phi_1$ is of order the decay constant 
$f$. After $t_1$, the number of ULALPs is conserved.   So their 
number density at later times is
\be 
n(t)\sim m \phi_1^2 \left(\frac{a(t_1)}{a(t)}\right)^3\,,
\label{eq:density}
\ee
where $a(t)$ is the cosmic scale factor. By demanding that the ULALPs 
make up the majority of CDM, we relate their initial number density to 
their mass,
\be 
n(t_1)\sim \frac{\rho_{c,0}}{m}\left(\frac{t_0}{t_{\rm eq}}\right)^2 
\left(\frac{t_{\rm eq}}{t_1}\right)^{3/2}\,,
\ee
where $t_0$ and $\rho_{{\rm c},0}$ are the age of the Universe and cold 
dark matter density today, respectively.  We assumed that $t_{1}$ is in 
the radiation-dominated era, which requires $m > 2 \times 10^{-28}$ eV.   We 
have therefore
\be 
\phi_1 \sim \frac{\sqrt{\rho_{{\rm c}, 0}}t_0}{(m t_{\rm eq})^{1 \over 4}}
\sim 3 \times 10^{17}~{\rm GeV} \left({10^{-22}~{\rm eV} \over m}\right)^{1 \over 4} 
 \,.
\ee
If inflation does not homogenize the ULALP field, the cold ULALPs produced by 
vacuum realignment have momentum dispersion of order 
\be 
\delta p(t)\sim \frac{1}{t_1}\left(\frac{a(t_1)}{a(t)}\right)\, ,\label{momdis}
\ee
and hence their average state occupation number is
\be 
{\mathcal{N}} \sim \frac{(2\pi)^3}{4\pi / 3}\frac{n(t)}{(\delta p)^3}
\sim 5 \times 10^{98} \left({10^{-22}~{\rm eV} \over m}\right)^{5 \over 2}
\,.
\label{eq:occ_n}
\ee
If inflation does homogenize the ULALP field, the momentum dispersion 
is smaller yet, and the average quantum state occupation number is higher.
Therefore, the ULALPs certainly form a highly degenerate Bose gas. 

The decay rate of ULALPs into two photons, in analogy with the QCD axion, 
is of order
\be 
\Gamma_{a\gamma\gamma}\sim
\frac{1}{64\pi}\Big(\frac{\alpha}{\pi}\Big)^2\frac{m^3}{f^2}
= {1 \over 2.5 \times 10^{110}~{\rm sec}} 
\left({m \over 10^{-22}~{\rm eV}}\right)^3
\left({10^{17}~{\rm GeV} \over f}\right)^2
\,
\ee
where $\alpha$ is the fine structure constant.  The ULALP is thus stable 
on the time scale of the age of the Universe.  More generally, because 
all ULALP number changing interactions are suppressed by one or more
powers of $1/f$, the number of ULALPs is conserved on time scales of 
order the age of the Universe. So, the third condition for ULALP 
Bose-Einstein condensation is satisfied as well.

The fourth condition is that ULALPs thermalize on a time scale 
shorter than the age of the universe.  We call the time scale 
over which the momentum distribution of the ULALPs changes 
completely as a result of their self-interactions the relaxation 
time $\tau$.  The relaxation rate is $\Gamma \equiv 1/\tau$.  
ULALPs certainly have gravitational self-interactions but 
perhaps also $\lambda \phi^4$ self-interactions.  If 
thermalization occurs, it occurs in a regime where the 
energy dispersion $\delta \omega$ is smaller than the relaxation 
rate $\Gamma$.   Indeed, since Eq.~(\ref{momdis}) gives an upper 
limit on the momentum dispersion, we have 
\begin{eqnarray} 
\delta \omega(t)  \sim {(\delta p(t))^2 \over 2m} 
< {1 \over 2 m t_1^2} \left({a(t_1) \over a(t)}\right)^2 
&=& H ~~~~~~~~~~~~~~~~~~{\rm for}~~ t < t_{\rm eq}\nonumber\\
&=& {3 \over 4} \left({t_{\rm eq} \over t}\right)^{1 \over 3} H
~~~~~{\rm for}~~ t > t_{\rm eq}
\label{endis}
\end{eqnarray}
and $H < \Gamma$ is necessary for thermalization to occur.  The 
condition $\delta \omega < \Gamma$ defines the ``condensed regime".

In the condensed regime, the relaxation rate due to $\lambda \phi^4$
self-interactions is of order \cite{CABEC,Erken}
\begin{equation}
\Gamma_\lambda \sim {|\lambda| n \over 4 m^2}~~~\ .
\label{lrel}
\end{equation}
For QCD axions, $|\lambda| \sim {m^2 \over f^2}$.  If we assume the 
same holds true for ULALPs and furthermore that $\phi_1 \sim f$, we
have 
\begin{eqnarray}
\Gamma_\lambda(t)/H(t) \sim 
{n(t_1) \over 2 f^2} \left({a(t_1) \over a(t)}\right)^3 t 
&\sim& \left({t_1 \over t}\right)^{1 \over 2}
~~~~~~~~{\rm for}~~~ t < t_{\rm eq} \nonumber\\
&\sim& {\sqrt{t_1 t_{\rm eq}} \over t}~~~~~~~~~{\rm for}~~~ t > t_{\rm eq}~~~\ .
\label{gloh}
\end{eqnarray}
Thus we find that, like QCD axions, the ULALPs may briefly thermalize 
through their $\lambda \phi^4$ self-interactions when they are first 
produced by vacuum realignment in the early Universe but that they 
will at any rate stop doing so shortly thereafter.

In the condensed regime, the relaxation rate due to gravitational 
self-interactions is of order \cite{CABEC,Erken}
\begin{equation}
\Gamma_{\rm g}\sim 4 \pi G n m^2 \ell^2
\label{grel}
\end{equation}
where $\ell \equiv 1/\delta p$ is the correlation length.  Since 
Eq.~(\ref{momdis}) gives an upper limit on the momentum dispersion, 
and we assume that ULALPs constitute most of the dark matter, we have 
\begin{eqnarray}
\Gamma_{\rm g}(t)/H(t) &\sim& {3 H \over 2} {n(t) m^2 \over \rho_{\rm tot}(t)} 
{1 \over (\delta p (t))^2} 
\gtrsim {3 H \over 2 m} {\rho(t) \over \rho_{\rm tot}(t)} 
\left({a(t) \over a(t_1)}\right)^2\nonumber\\
&\gtrsim& \sqrt{t \over t_{\rm eq}}~~~~~~~~{\rm for}~~~~t < t_{\rm eq}\nonumber\\
&\gtrsim& \left({t \over t_{\rm eq}}\right)^{1 \over 3}
~~~~~{\rm for}~~~~t > t_{\rm eq}~~\ ,
\label{ggoh}
\end{eqnarray}
where $\rho = nm$.  We used the Friedmann equation to relate the total density 
$\rho_{\rm tot}$ to the Hubble constant.  Equation (\ref{ggoh}) shows that, independently 
of the ULALP mass, the ULALP fluid thermalizes by gravitational self-interactions 
at the time of equality, or earlier if the ULALP field was homogenized by 
inflation.
 
It was shown in Ref. \cite{Erken} that QCD axions that are about to fall into 
a galactic halo thermalize sufficiently fast by gravitational self-interactions
that they almost all go the lowest energy state available to them consistent 
with the total angular momentum they acquired from tidal torquing by galactic 
neighbors.   That lowest energy state is one of rigid rotation on the turnaround 
sphere.  It was shown in Ref.~\cite{case} that this redistribution of angular 
momentum among infalling dark matter axions explains precisely and in all 
respects the properties of caustic rings of dark matter for which observational 
evidence had been found earlier.   The observational evidence for caustic rings
of dark matter in the Milky Way and other isolated disk galaxies is summarized 
in Ref. \cite{Duffy}.  Furthermore, it was shown in Ref. \cite{Banik} that this 
redistribution of angular momentum solves the galactic angular momentum problem, 
which is the tendency of cold dark matter, in numerical simulations of structure
formation, to be too concentrated at galactic centers.  The fact that Bose-Einstein 
condensation of the dark matter particles explains the observational evidence for 
caustic rings and solves the galactic angular momentum problem and the fact that 
Bose-Einstein condensation is a property of dark matter axions but not of the 
other dark matter candidates, constitute an argument that the dark matter is 
axions, at least in part \cite{argu}.  Although these studies \cite{Erken,case,Banik} 
were motivated primarily by QCD axions, they do not depend in an essential way on the 
axion mass and apply equally well to any axionlike dark matter candidate produced by 
the vacuum realignment mechanism, including ULALPs.  To summarize, if ULALPs are 
the dark matter, they thermalize by gravitational interactions and form 
a Bose-Einstein condensate at or before the time of equality between matter 
and radiation. They thermalize sufficiently fast before falling onto galactic 
halos to acquire quasirigid rotation on the turnaround sphere.  They then
form caustic rings with properties that are in accord with observations.  The 
observational evidence for caustic rings therefore supports the hypothesis 
that the dark matter is ULALPs but also, as we will see, constrains that 
hypothesis.

We conclude this section with a discussion of the results of Ref. \cite{Hui}
on ULALP relaxation in gravitationally bound objects.  We will show that the 
relaxation rate obtained there is consistent with the earlier results on axion 
relaxation in Refs. \cite{CABEC,Erken}.  The relaxation rate in the particle 
kinetic regime is \cite{Erken}
\begin{equation}
\Gamma \sim n~\sigma~\delta v~{\cal N}
\label{pkrr}
\end{equation}
where $n$ is the particle density, $\sigma$ is the scattering cross section, 
$\delta v$ is the velocity dispersion, and ${\cal N}$ is the degeneracy (i.e.,the average occupation number of those particle states that are occupied).
The relevant cross-section for relaxation by gravitational interactions in 
the particle kinetic regime is \cite{Erken}
\begin{equation}
\sigma_g \sim {4 G^2 m^2 \over (\delta v)^4}~~\ .
\label{gxc}
\end{equation}
The average quantum degeneracy is
\begin{equation}
{\cal N} \sim {(2 \pi)^3 n \over {4 \pi \over 3} (m \delta v)^3}~~\ .
\label{aqd}
\end{equation}
Combining Eqs. (\ref{pkrr} - \ref{aqd}) gives the relaxation rate by 
gravitational interactions in the particle kinetic regime
\begin{equation}
\Gamma_g \sim 24 \pi^2 {(G \rho)^2 \over m^3 (\delta v)^6}~~\ .
\label{genrel} 
\end{equation}
For the special case of a gravitationally bound object, the crossing 
time is of order $t_{\rm cr} \sim \sqrt{3 \pi \over 4 G \rho}$ and 
the size of the object is of order $r \sim \delta v t_{\rm cr}/\pi$.
The relaxation time is then of order
\begin{equation}
\tau_g = {1 \over \Gamma_g} \sim {2 \over 27} r^4 m^3 (\delta v)^2 ~~\ . 
\label{relti}
\end{equation}
Ignoring the numerical prefactor, which is at any rate poorly known, 
this is the result given in Ref. \cite{Hui} for gravitationally bound 
systems.  Hui \textit{et al.} consider several systems that may relax in the 
particle kinetic regime and find either that they do not relax or that 
relaxation has no observable consequences.  Whether the ULALP fluid 
relaxes or not is of paramount importance in determining its behavior 
because it does not obey classical field equations when it relaxes.  We 
found earlier in this section, following the discussion in ref. \cite{Erken}, 
that the ULALP fluid does relax before it falls into galactic halos.  It was 
our purpose in this paragraph to show that the general discussion of cosmic 
axion relaxation in Ref. \cite{Erken} is consistent with both the relaxation 
rate for gravitationally bound objects given in ref. \cite{Hui} and our 
claim that the ULALP fluid relaxes before falling into galactic halos. 

\section{Bound from ULALP infall}

Provided no thermalization is occurring \cite{Erken,LNpt}, ULALP dark 
matter is represented by a wave function solving the Schr\"odinger-Poisson 
equations.   The wave function may be written in terms of a real amplitude 
$A$ and a phase $\beta$:
\begin{equation}
\Psi(\vec{r},t) = A(\vec{r},t) e^{i \beta(\vec{r},t)}~~\ .
\label{wvfct}
\end{equation}
In the linear regime of the growth of density perturbations, before 
multistreaming occurs, Eq.~(\ref{wvfct}) describes a flow of density,
\begin{equation}
n(\vec{r},t) = (A(\vec{r},t))^2
\label{den}
\end{equation}
and velocity,
\begin{equation}
\vec{v}(\vec{r},t) = {1 \over m} \vec{\nabla} \beta(\vec{r},t)~~\ .
\label{vel}
\end{equation}
In particular, a homogeneous flow of density $n$ and velocity $\vec{v}$
is described by
\begin{equation}
\Psi(\vec{r},t) = \sqrt{n}~e^{i (\vec{p} \cdot\vec{r} - \omega t)},
\label{hom}
\end{equation}
where $\vec{p} = m \vec{v}$ is the momentum and $\omega = p^2/2m$ is the 
energy of each particle. An attractive feature of the wave function 
description is that it readily accommodates multistreaming \cite{KW}.  
In a region with two homogeneous cold dark matter flows, with densities 
$n_i$ and velocities $\vec{v}_i~(i=1,2)$, the wave function is
\begin{equation}
\Psi(\vec{r},t) = \sqrt{n_1}~e^{i (\vec{p}_1\cdot\vec{r} - \omega_1 t)}
+ \sqrt{n_2}~e^{i (\vec{p}_2 \cdot\vec{r} - \omega_2 t - \delta)}~~\ .
\label{2fl}
\end{equation}
The associated density is
\begin{equation}
n(\vec{r},t) = |\Psi(\vec{r},t)|^2 = n_1 + n_2 
+ 2 \sqrt{n_1 n_2} \cos(\Delta\vec{p}\cdot\vec{r} 
- \Delta \omega t - \delta)~~\ ,
\label{2flden}
\end{equation}
where $\Delta \vec{p} = \vec{p}_2 - \vec{p}_1$ and 
$\Delta \omega = \omega_2 - \omega_1$.  The interference 
term is important for ULALP dark matter because the  
correlation length $\ell = {1 \over |\Delta\vec{p}|}$ 
is large (since $m$ is small and $\Delta\vec{v}$ is generally 
fixed by observation) and gravity is long range.  The 
gravitational field sourced by the interference term in 
Eq.~(\ref{2flden}) is 
\begin{equation}
\vec{g} = - 8 \pi G \sqrt{n_1 n_2}~ m \ell~ 
\sin(\Delta \vec{p}\cdot\vec{r} - \Delta \omega t - \delta)
~\hat{n}
\label{2flgr}
\end{equation}
where $\hat{n} = \ell \Delta \vec{p}$.  It is the gravitational 
fields associated with the interference terms in the ULALP fluid 
density that cause the ULALP dark matter thermalization discussed 
in the previous section \cite{Erken}.
 
An isolated galaxy such as our own Milky Way continually accretes surrounding 
dark matter.   There are on Earth two flows of dark matter falling in and out 
of the Galaxy for the first time ($n$ =1), two flows falling in and out for the 
second time ($n$ = 2), two flows falling in and out for the third time ($n$ = 3), 
and so forth.  The total number of flows on Earth is of order the age of the 
Universe divided by the time it takes a particle initially at rest at our 
galactocentric distance to fall through the Galaxy and reach the opposite 
side, i.e., $10^{10}~{\rm yr}/10^8~{\rm yr} = 100$.  The flows of 
particles that fell into the halo relatively recently ($n \lesssim 10$) 
have not been thermalized as a result of gravitational scattering off 
inhomogeneities in the Galaxy \cite{Ipser}. Those flows and their associated 
caustics are a robust prediction \cite{Natarajan} of galactic halo formation 
with cold dark matter.

For $n \lesssim 10$, the density of each flow is of order 2\% or 3\% 
\cite{Ipser,STW,Duffy} of the total average dark matter density 
at our distance from the Galactic center; i.e., each flow has density of 
order $\rho_{\rm fl} \sim 10^{-26}$ gr/cm$^3$.  Since the typical difference 
in velocities between pairs of flows is of order $10^{-3}c$, the coherence 
lengths associated with the interference terms are of order
\begin{equation}
\ell \sim {1 \over 10^{-3} m} 
\sim 64~{\rm pc}~\left({10^{-22}~{\rm eV} \over m}\right)
\label{corl}
\end{equation} 
and their correlation times are of order
\begin{equation}
T \sim {1 \over 10^{-6} m} 
\sim 2 \times 10^5~{\rm year} ~\left({10^{-22}~{\rm eV} \over m}\right)~~\ .
\label{cort}
\end{equation}
The gravitational fields sourced by the interference terms are of order  
\begin{equation}
g \sim 4 \pi G \rho_{\rm fl}~\ell
\sim 0.5~{{\rm km} \over {\rm sec} \times {\rm Gyear}}~
\left({10^{-22}~{\rm eV} \over m}\right)~~\ .
\label{gravacc}
\end{equation}
A star in the galactic disk acquires from each pair of flows
a velocity increment of order $g T$ after a coherence time $T$.  
The total velocity acquired by a star is the result of a random 
walk in its velocity space, and of order
\begin{equation}
\Delta v \sim gT \sqrt{t_0 \over T} \sqrt{{1 \over 2} N(N-1)}
\sim 0.5~{{\rm km} \over {\rm sec}} 
\left({10^{-22}~{\rm eV} \over m}\right)^{3 \over 2}
\label{starvel}
\end{equation}
where $t_0 \sim 10^{10}$ yr is the age of the Universe and 
$N \sim 20$ is the number of flows with density $\rho_{\rm fl}$. 
Note that $\Delta v$ is not sharply sensitive to our assumption 
on the number of flows since $\rho_{\rm fl}$ and $g$ are inversely 
proportional to $N$, whereas the number of flow pairs is proportional 
to $N^2$.  The thin disk of the Milky Way is made of stars with 
vertical velocity dispersion $\sigma_z \sim 20$ km/s$^{-1}$.  
We obtain a lower bound on the ULALP mass
\begin{equation}
m \gtrsim 10^{-23}~{\rm eV}
\label{bound}
\end{equation}
by requiring $\Delta v < \sigma_z$.  The result obtained here is 
broadly consistent with the discussion of the thickening of the 
Galactic disk by ULALP dark matter in Ref. \cite{Hui}.

It was observed in Ref.~\cite{Khmelnitsky2014} that ULALP dark matter 
has an oscillating pressure/tension with angular frequency equal to 
twice the ULALP mass and that this oscillating pressure/tension, being 
a source of gravity in general relativity, has measurable effects.  The 
idea presented here is similar but makes use of Newtonian gravity and 
multistreaming instead.  

\section{Bound from the sharpness of the nearby caustic ring}

Cold, collisionless dark matter lies in six-dimensional phase space 
on a thin three-dimensional hypersurface, the thickness of which is the 
velocity dispersion of the dark matter particles.  We refer to this 
hypersurface as the ``phase space sheet".  As dark matter particles 
fall in and out of a galactic gravitational potential well, their 
phase space sheet wraps up. Locations where the phase space sheet 
folds back onto itself have large particle density in physical space 
\cite{STW}, \cite{Sikivie:1997ng}, \cite{singul}, \cite{Natarajan}. 
Indeed, the phase space sheet is tangent to velocity space at the 
location of such folds, and therefore the physical space density 
diverges there in the limit of vanishing sheet thickness.  For 
small velocity dispersion, the physical density at the location 
of a fold is finite but very large.  These locations of high density 
are called caustics.  They are generically two-dimensional surfaces in 
physical space.  The kinds of caustics that appear are classified by 
catastrophe theory.

It was shown \cite{Natarajan} that galactic halos built up by 
the infall of cold collisionless dark matter have two sets of 
caustics, inner and outer.  There is one inner and one outer 
caustic for each value of the integer $n$ introduced in the 
previous section.  The catastrophe structure of the inner 
caustics depends on the angular momentum distribution of the 
infalling particles.  If the velocity field of the infalling 
particles is dominated by net overall rotation 
($\vec{\nabla} \times \vec{v} \neq 0$), the inner caustics are 
closed tubes of which the cross section is a section of the elliptic 
umbilic ($D_{-4}$) catastrophe \cite{singul}, called caustic 
rings for short.  Evidence for caustic rings of dark matter 
was derived from a variety of observations.  The evidence is 
summarized in Ref.~\cite{Duffy}.  The evidence is explained in 
all its aspects if the dark matter is axions or axionlike 
particles such as ULALPs \cite{case}.  

The caustic ring model of the Milky Way halo was independently 
tested against observations in Ref. \cite{Newberg}.  Recently, 
Y. Huang \textit{et al.} \cite{Huang} found evidence for the second and 
third ($n$ = 2,3) caustic rings in our Galaxy from their measurement 
of the Milky Way rotation curve out to very large (100 kpc) radii.  

\begin{figure}[h!]
\includegraphics[scale=.7]{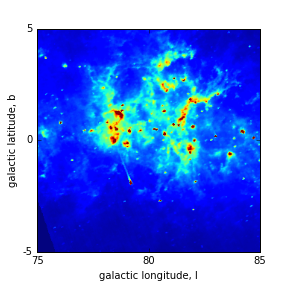}
\includegraphics[scale=.7]{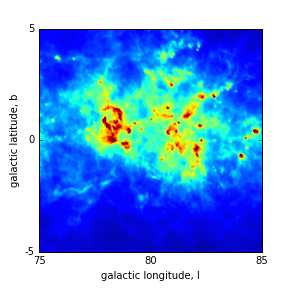}
\caption{Images of the region around galactic coordinates $(80,0)$ from 
the IRAS 12 $\mu {\rm m}$ and Planck {857~GHz} observations. The triangular 
feature, indicative of a tricusp caustic ring, is prominent in both data sets.}
\label{fig:tri}
\end{figure}

\begin{figure}[h!]
\includegraphics[scale=.14]{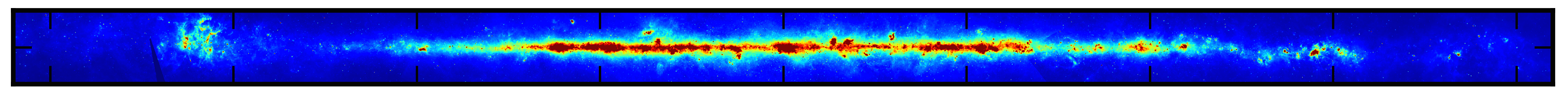}
\caption{Panoramic view of the Milky Way Galactic plane from the IRAS experiment in 
the $12~ \rm{ \mu m}$ band. }
\label{fig:pana}
\end{figure}

Caustic rings of dark matter produce sharp rises in galactic 
rotation curves \cite{singul}.  Part of the observational 
evidence for caustic rings is that sharp rises appear in the 
Milky Way rotation curve \cite{Clemens} at radii consistent 
with the predictions for caustic ring radii by the self-similar 
infall model \cite{MWcr}.  Furthermore, there is a triangular 
feature in the IRAS map \cite{Neugebauer:1984zz} of the Galactic 
plane of which position on the sky matches the position of the sharp 
rise in the Milky Way rotation attributed to the caustic ring 
($n = 5$) nearest to us.  The triangular feature appears in the 
direction of galactic coordinates $(l,b)$ = (80,0). It is seen 
also in the recent Planck observations \cite{Planck2015}; see 
Fig.~\ref{fig:tri}. In Fig.~\ref{fig:pana} we show the panoramic 
view of the Milky Way provided by the IRAS $12~\rm{\mu m}$ observations. 
The triangular feature is clearly visible on the left-hand side 
of the image.  It can be understood as the imprint of the nearby 
caustic ring of dark matter upon the gas and dust in the Galactic 
disk \cite{MWcr}.

The sharpness of the triangular feature implies that its edges 
are not smoothed on distances larger than approximately 10 pc.  
Because velocity dispersion smooths caustics, the sharpness of 
the triangular feature implies an upper limit of order 50 m/s 
\cite{MWcr,veldis} on the velocity dispersion of the flow ($n = 5$) 
that produces the nearby caustic ring.   As we show now, it also 
implies a lower limit on the mass of ULALP dark matter, because 
the wave description smooths caustics as well.   

Consider the simple fold caustic that occurs at a caustic ring 
radius, where the particles that fall in along the galactic 
plane reach their distance of closest approach to the galactic 
center before falling back out.  In the particle description, the
particles satisfy the equation of motion
\begin{equation}
m {d^2 r \over dt^2} = - {d V_{\rm eff}(r) \over dr}
\label{Newton}
\end{equation} 
where $r$ is the distance to the galactic center and $V_{\rm eff}(r)$ 
is the effective potential for radial motion, including the repulsive 
angular momentum barrier.  In the neighborhood of the caustic, the 
radial velocity of the particles is 
\begin{equation}
v_r(r) = \pm \sqrt{ - {2 \over m} {d V_{\rm eff} \over dr}(a) (r - a)}
\label{vr}
\end{equation}
where $a$ is the caustic ring radius.  Conservation of the number 
of particles implies that the density near the caustic is proportional 
to
\begin{equation}
n(r) \propto {1 \over r^2 |v_r(r)|}~~~\ .
\label{cden}
\end{equation}
The density diverges therefore at $r=a$ as $1/\sqrt{r-a}$, which is 
the characteristic behavior of a simple fold caustic.   Spreading of 
the caustic due to velocity dispersion is discussed in Ref. \cite{veldis}.

In the wave description, the behavior of the dark matter fluid is 
given by a wavefunction $\Psi(\vec{r})$ that satisfies the Schr\"odinger
equation.   At the location of the fold discussed in the previous paragraph,
the radial part of the wave function satisfies
\begin{equation}
- {\hbar^2 \over 2 m} {1 \over r^2} {d \over dr} r^2 {d \Psi(r) \over dr}
+ V_{\rm eff}(r) \Psi(r) = E \Psi(r) 
\label{schro}
\end{equation}
with $E = V_{\rm eff}(a)$.  Near $r = a$, the wave function is proportional 
to an Airy function \cite{Griffiths}
\begin{equation}
\Psi(r) = C {1 \over r} Ai(\gamma (a - r))
\label{Air}
\end{equation}
with 
\begin{equation}
\gamma = 
\left[- {2 m \over \hbar^2} {d V_{\rm eff} \over dr}(a) \right]^{1 \over 3}~~~\ .
\label{gam}
\end{equation}
The wave function smooths the fold caustic over a distance scale $\gamma^{-1}$.
The sharpness of the triangular feature in the IRAS map associated with the 
fifth caustic ring implies an upper bound on $\gamma^{-1}$ of order 10 pc. The 
dominant contribution to $ - {d V_{\rm eff} \over dr}(a)$ is the centrifugal 
force $m v_\phi(a)^2/a$ where $v_\phi(a) \simeq 500$ km/s is the velocity 
component of the flow producing the fifth caustic ring in the direction of 
galactic rotation.  We obtain the bound 
\begin{equation}
m \gtrsim 10^{-20}~{\rm eV}
\label{2b}
\end{equation}
by requiring $\gamma^{-1} \lesssim$ 10 pc.  Finally we note that high- 
resolution observation of the nearby caustic provides, in principle, a 
means to determine the ULALP mass since the gravitational field in the 
plane of the ring near $r=a$ has characteristic structure on the length 
scale $\gamma^{-1} \propto m^{-2/3}$.

\section{Summary}


In this paper, we have explored ULALPs, an alternative cold dark matter 
candidate discussed by many authors. We showed that ULALPs thermalize 
through gravitational self-interactions and form a Bose-Einstein condensate 
in a manner analogous to QCD axions. This was not taken into account in the 
previous literature.

We placed new constraints on the mass of ULALPs. First, we considered 
the heating effect of infalling ULALPs on the disk stars of the Milky 
Way galaxy. We derived therefrom a lower bound of the ULALP mass, which 
is comparable in strength to the previous bounds from structure formation 
and cosmic microwave background data. 

Our tightest bound comes from taking into account the Bose-Einstein 
condensate nature of ULALPs. In a way analogous to the QCD axion, 
Bose-Einstein condensation enables the ULALP dark matter to acquire 
net overall rotation and hence form caustic rings with the tricusp cross 
section. From the observed sharpness of the triangular feature due to 
the fifth caustic ring in the Milky Way, we derive a bound on the spread 
of the ULALP wave function and, in turn, on the ULALP mass. The bound we 
obtain, $m_a\gtrsim 10^{-20}\, {\rm eV}$, disfavors a large fraction of 
the ULALPs from being the dark matter. It is in tension with recent 
works that find $m \sim 10^{-22}\, {\rm eV}$ fits the data best.

\section*{Acknowledgements}
P. S. would like to thank Qiaoli Yang for early discussions on this topic. At the 
University of Florida this work is supported in part by the U.S. Department of 
Energy under Grant No. DE-FG02-97ER41029 and by the Heising-Simons Foundation 
under Grant No. 2015-109.  Fermilab is operated by Fermi Research Alliance, LLC, 
under Contract No. DE-AC02-07CH11359 with the U.S. Department of Energy. N. B. was 
supported by the Fermilab Graduate Student Research Program in Theoretical 
Physics.


\begin{thebibliography}{bib}

\bibitem{Peccei1977}
R.~Peccei, and H.~R. Quinn, Phys. Rev. Lett. 38 (1977) 1440.

\bibitem{Weinberg1978}
S.~Weinberg, Phys. Rev. Lett. 40 (1978) 223.
  
\bibitem{Wilczek1978}
F.~Wilczek, Phys. Rev. Lett. 40 (1978) 279.
  
\bibitem{Preskill1983}
J.~Preskill, M.~B. Wise and F. Wilczek, Phys. Lett. B120 (1983) 127.

\bibitem{Abbott1983}
L.~Abbott and P.~Sikivie, Phys. Lett. B120 (1983) 133.
 
\bibitem{Dine1983}
M.~Dine and W.~Fischler, Phys. Lett. B120 (1983) 137.

\bibitem{Svrcek:2006yi}
P.~Svrcek and E.~Witten, JHEP 06 (2006) 051.

\bibitem{Arvanitaki:2009fg}
A.~Arvanitaki et al., Phys. Rev. D81 (2010) 123530.

\bibitem{Ringwald:2012hr}
A.~Ringwald, Phys. Dark Univ. 1 (2012) 116.

\bibitem{Press}
W.H. Press, B.S. Ryden and D.N. Spergel, Phys. Rev. Lett. 64 (1990) 1084.

\bibitem{Sin}
S.-J. Sin, Phys. Rev. D50 (1994) 3650.

\bibitem{Goodman}
J. Goodman, New Astronomy Reviews 5 (2000) 103.

\bibitem{Hu2000}
W.~Hu, R.~Barkana and A.~Gruzinov, Phys. Rev. Lett. 85 (2000) 1158.

\bibitem{Amendola:2005ad}
L.~Amendola and R.~Barbieri, Phys. Lett. B642 (2006) 192.

\bibitem{BoylanKolchin:2011de}
M.~Boylan-Kolchin, J.~Bullock and M.~Kaplinghat, MNRAS 415 (2011) L40.

\bibitem{Shapiro}
T. Rindler-Daller and P. Shapiro, MNRAS 422 (2012) 135.  

\bibitem{Park:2012ru}
C.-G.~Park, J.-c.~Hwang and H.~Noh, Phys. Rev. D86 (2012) 083535.

\bibitem{Khlopov}
M.Y. Khlopov, B.A. Malomed and Y.B. Zeldovich,               
MNRAS 215 (1985) 575.

\bibitem{Bianchi}
M. Bianchi, D. Grasso and R. Ruffini, Astron. Astrophys. 231
(1990) 301.

\bibitem{Lora}
V. Lora, J. Maga\~na, A. Bernal, F.J. S\'anchez-Salcedo and E.K. Grebel,
JCAP 2 (2012) 011.

\bibitem{Hlozek:2014lca}
R.~Hlozek, D.~Grin, D.~J.~Marsh and P.~G.~Ferreira, Phys. Rev. D91 (2015) 103512.

\bibitem{Nature}
H.Y. Schive, T. Chiueh and T. Broadhurst, Nature Phys. 10 (2014) 496.
  
\bibitem{Schive:2015kza}
H.-Y.~Schive, T.~Chiueh, T.~Broadhurst and K.-W. Huang, Ap. J. 818 (2016) 89.

\bibitem{MS}
D.J.E. Marsh and J. Silk, MNRAS 437 (2014) 2652.

\bibitem{Calabrese:2016aa}
E.~Calabrese and D.~N.Spergel, MNRAS 460 (2016) 4397.

\bibitem{Gonzales}
A.X. Gonz\'{a}les-Morales, D.J.E. Marsh, J. Pe\~{n}arrubia and L. Ure\~{n}a-L\'{o}pez, arXiv:1609.05856.

\bibitem{Bozek2015}
B.~Bozek, D.~J.~E.~Marsh, J.~Silk and R.~F.~G. Wyse, MNRAS 450 (2015) 209.

\bibitem{Hui}
L. Hui, J.P. Ostriker, S. Tremaine and E. Witten, arXiv:1610.08297.

\bibitem{bhupal}
P. S. Bhupal Dev, M. Lindner, and S. Ohmer, arXiv:1609.03939 

\bibitem{CABEC}
P. Sikivie and Q. Yang, Phys. Rev. Lett. 103 (2009) 111301.

\bibitem{Erken}
O. Erken, P. Sikivie, H. Tam and Q. Yang, 
Phys. Rev. D85 (2012) 063520.

\bibitem{Saikawa}
K. Saikawa and M. Yamaguchi, Phys. Rev. D87 (2013) 085010.

\bibitem{Berges}
J. Berges and J. Jaeckel, Phys. Rev. D91 (2015) 025020.

\bibitem{case}
P. Sikivie, Phys. Lett. B695 (2011) 22.

\bibitem{Duffy}
L. Duffy and P. Sikivie, Phys. Rev. D78 (2008) 063508.

\bibitem{Banik}
N. Banik and P. Sikivie, Phys. Rev. D88 (2013) 123517.

\bibitem{argu}
P. Sikivie, Springer Proc. Phys. 148 (2013) 25.

\bibitem{LNpt}
N. Banik, A. Christopherson, P. Sikivie and E. Todarello, 
Phys. Rev. D91 (2015) 123540.

\bibitem{KW}
L. Widrow and N. Kaiser, Ap. J. 416 (1993) L71.

\bibitem{Ipser}
P. Sikivie and J. Ipser, Phys. Lett. B291 (1992) 288.

\bibitem{Natarajan}
A. Natarajan and P. Sikivie, Phys. Rev. D73 (2006) 023510.

\bibitem{STW}
P. Sikivie, I. Tkachev and Y. Wang, Phys. Rev. Lett 75 (1995) 2911 
and Phys. Rev. D56 (1997) 1863.

\bibitem{Khmelnitsky2014}
A.~Khmelnitsky and V.~Rubakov, JCAP 1402 (2014) 019.
  
\bibitem{Sikivie:1997ng}
P.~Sikivie, Phys. Lett. B432 (1998) 139.

\bibitem{singul}
P. Sikivie, Phys. Rev. D60 (1999) 063501.

\bibitem{Newberg}
J. Dumas et al., Ap. J. 811 (2015) 36.

\bibitem{Huang}
Y. Huang et al., MNRAS 463 (2016) 2623.

\bibitem{Clemens}
D.P. Clemens, Ap. J. 295 (1985) 422.

\bibitem{MWcr}
P. Sikivie, Phys. Lett. B567 (2003) 1.

\bibitem{Neugebauer:1984zz}
G.~Neugebauer et~al., Ap. J. 278 (1984) L1.

\bibitem{Planck2015}
Planck Collaboration, R. Adam et al., Astron.Astrophys. 594 (2016) A1. 

\bibitem{veldis}
N. Banik and P. Sikivie, Phys. Rev. D93 (2016) 103509.

\bibitem{Griffiths}
D.J. Griffiths, {\it Introduction to Quantum Mechanics}, 
Pearson Prentice Hall 2005.


\end{thebibliography}

\end{document}